\newcommand{\bra}[1]{\langle #1|}
\newcommand{\ket}[1]{|#1\rangle}

\documentclass[showpacs,pra,twocolumn,amsmath]{revtex4}
\usepackage{graphicx}
\usepackage{amsfonts}

\begin{document}
\title{Effects of environmental parameters to total, quantum and classical correlations}

\author{Wen-Ling Chan$^1$}
\author{Jun-Peng Cao$^{1,2}$}
\author{Dong Yang$^{1,3}$}
\author{Shi-Jian Gu$^1$}
\altaffiliation{Email: sjgu@phy.cuhk.edu.hk\\URL: http://www.phystar.net/}
\affiliation{$^1$Department of Physics and Institute of Theoretical Physics,
The Chinese University of Hong Kong, Hong Kong, China\\
$^2$Institute of
Physics, Chinese Academy of Sciences, Beijing 100080, China\\
$^3$Department of Physics, Hong Kong University, Hong Kong, China}

\begin{abstract}
We quantify the total, quantum, and classical correlations with entropic
measures, and quantitatively compare these correlations in a quantum system, as
exemplified by a Heisenberg dimer which is subjected to the change of
environmental parameters: temperature and nonuniform external field. Our
results show that the quantum correlation may exceed the classical correlation
at some nonzero temperatures, though the former is rather fragile than the
later under thermal fluctuation. The effect of the external field to the
classical correlation is quite different from the quantum correlation.
\end{abstract}

\date{\today}
\maketitle

\section{Introduction}

Correlation effect plays an important role in physical phenomena. Many
interesting properties of the quantum systems are attributed to the existence of
the entanglement \cite{EPR}, which is intrinsically related to the superposition
principle of quantum mechanics and the direct product structure of the Hilbert
space \cite{Nielsen1,AGalindo02}. Entanglement is a kind of pure quantum
correlation which does not exist in any classical systems, and regarded as a
significant resource in quantum information processing, such as quantum
teleportation, dense coding, and quantum cryptography \cite{Bennett}.

Due to the central role of the entanglement in quantum information, various
issues on the entanglement have been studied intensively in recent fifteen
years. Among these issue, the effects of environmental parameters (such as
thermal fluctuation and external field) to the entanglement in quantum spin
systems have been attracted much attention
\cite{XWang01,PZanardi,XQXi02,CAnteneodo03,YSun03}. Moreover, some interesting
properties of the entanglement which is beyond the traditional physical
intuition were found. For examples, the thermal fluctuation can enhance the
entanglement in some special cases, the external field does not always suppress
the entanglement \cite{YSun03}. Therefore, these studies shed new light on our
understanding of the entanglement.

However, besides the quantum entanglement, a quantum system possesses the
classical correlation \cite{HV,BGroisman05}. A simple example is the spin
singlet state $(\ket{\uparrow\downarrow}-\ket{\downarrow\uparrow})/\sqrt{2}$,
besides the entanglement with value 1, the state also has the classical
correlation of the value 1 (see section II). Another example is the mixed state
with the density matrix as
$\rho=(|\uparrow\uparrow\rangle\langle\uparrow\uparrow|
+|\downarrow\downarrow\rangle\langle\downarrow\downarrow|)/2$. In this state,
the quantum correlation between two spins is zero, while the classical
correlation is 1. Therefore, some interesting questions arise. For examples,
What is the difference between the classical correlation and quantum correlation
in a realistic system? Is the classical correlation always larger than the
quantum one? Why we live in a classical world rather than a quantum world? etc..
Answering these questions from the point of view of different correlations are
our main motivations in this work.

Our paper is organized as follows. In section \ref{sec:definition}, we define
the measurements of total, quantum and classical correlations. In section
\ref{sec:enviroment}, we use a Heisenberg dimer which interacts with the thermal
environment as an example to study the effects of the temperature, external
fields, and anisotropic interaction on the the correlations. Finally, a summary
is given in section \ref{sec:sum}.

\section{Definitions and measures of bipartite correlations}
\label{sec:definition}

\subsection{Total correlation}
In quantum information theory \cite{Nielsen1}, for two subsystems 1 and 2, the
mutual information is defined as
\begin{equation}
S(1:2)=S(1)+S(2)-S(1\cup 2), \label{eq:S(1:2)}
\end{equation}
where $S(i)=-{\rm tr}(\rho_i\,{\rm log}_2\,\rho_i),i=1,2,1\cup 2$ is the entropy
of the corresponding reduced density matrix. Since the entropy is used to
quantify the physical resource (in unit of classical bit due to $\log_2$ in its
expression) needed to store information of a system, the mutual entropy then
measures additional physical resource required if we store two subsystems
respectively rather than store them together. Let us look at a very simple
example: a two-qubit system in a singlet state
$(\ket{\uparrow\downarrow}-\ket{\downarrow\uparrow})/\sqrt{2}$. We have
$S(1)=S(2)=1$ and $S(1\cup 2)=0$, hence $S(1:2)=2$. Obviously, there is no
information in a given singlet state. However, each spin in this state is
completely uncertain. So we need two bits to store them respectively. Here the
mutual information is twice the entanglement, as measured by the von Neumann
entropy of either subsystem. This is due to the reason that besides quantum
correlation, the state has also classical correlation between the two
subsystems. Therefore, the mutual information can be used to measure the total
correlation between two subsystems. We will call the quanity $S(1:2)$ as the
``total correlation entropy'' or simply ``total correlation'' hereafter.

\subsection{Quantum correlation}

The quantum correlation only exists in the quantum world, and usually is called
entanglement. For bipartite state, there are a few measures to quantify the
entanglement of a general mixed state \cite{PV}. Among the measures, the
entanglement of formation \cite{CHBennett96} is well known and a analytic
formula for two-qubit system is found \cite{SHill97}. Consider a density matrix
$\rho$ of two subsystems 1 and 2. There are infinite pure-state ensembles
$\{\psi_i,p_i\}$ of $\rho$, where $p_i$ is the probability of $\psi_i$, such
that
\begin{equation}
\rho=\sum_i p_i \ket{\psi_i}\bra{\psi_i}. \label{decomposition}
\end{equation}
For each pure state $\ket{\psi_i}$, the entanglement $E$ is measured by the von
Neumann entropy \cite{AWehrl94}. Then the entanglement of formation $E_f$ of the
density matrix $\rho$ is the average entanglement of the pure states of the
decomposition, minimized over all the possible ensembles:
\begin{equation}
E_f(\rho)={\rm min}\sum_i p_i E(\psi_i). \label{eq:Ef}
\end{equation}
(Note that if the system is in a pure state, $E_f$ is just $E$.) For a mixed
state, it is usually difficult to evaluate $E_f$. However, for a two-qubit
system, it can be readily obtained from the concurrence of the system. Given the
density matrix $\rho$ of the pair qubits, the concurrence is given by
\cite{SHill97}
\begin{equation}
C={\rm max}\{\lambda_1-\lambda_2-\lambda_3-\lambda_4,0\}, \label{eq:con}
\end{equation}
where $\lambda_i$ are the square roots of the eigenvalues of the operator
\begin{equation}
\varrho=\rho(\sigma_1^y\otimes\sigma_2^y)\rho^*(\sigma_1^y\otimes\sigma_2^y),
\label{eq:varrho}
\end{equation}
with $\lambda_1\geq\lambda_2\geq\lambda_3\geq\lambda_4$, $\sigma_i^y$ are the
normal Pauli operators, and $\rho^*$ is the complex conjugate operator of
$\rho$. The entanglement of formation can then be evaluated as \cite{SHill97}
\begin{eqnarray}
&& E_f=h\left({1+\sqrt{1-C^2}}\over2\right); \nonumber \\
&& h(x)=-x\,{\rm log}_2 \, x-(1-x)\,{\rm log}_2 \, (1-x). \label{eq:Eof}
\end{eqnarray}
$E_f$ is monotonically increasing and ranges from 0 to 1 as $C$ goes from 0 to
1. $E_f=C=0$ if the system is unentangled and $E_f=C=1$ if it is maximally
entangled. In fact, one can take the concurrence itself as a measurement of
entanglement. Since the mutual information has the unit of bit, for comparison
purpose, we will take the entanglement of formation instead of concurrence to be
our measurement standard in this paper. We will call the quantity $E_f$ as the
``quantum correlation entropy'' or simply ``quantum correlation'' hereafter.

\subsection{Classical correlation}
The classical correlation of a bipartite system is defined in different
scenarios \cite{HV,BGroisman05}. The measure defined in \cite{HV} reflects the
effect of one party's measurement on the other party's state. The measure
defined in \cite{BGroisman05} attempts to explain the total correlation coming
from quantum part and classical part based on the distance concept of relative
entropy. Both these two measure coincides in the case of pure states. Consider a
pure state of bipartite system,
$\ket{\psi}=\sum_i\alpha_i\ket{u_i}\otimes\ket{v_i}$ unpon Schmidt
decomposition. The quantum correlation actually defines the amount of immediate
effect on one subsystem during the performing measurement on another subsystem.
For pure state, it is just the entropy of one subsystem. After the measurement,
the density matrix becomes diagonal in the basis of the Schmidt decomposition.
Then the classical correlation between these two subsystems corresponds to the
maximum amount of change of uncertainty in one subsystem after knowing some new
information of another subsystem through a classical channel. Such a correlation
equals to the entropy of one subsystems too. For the mixed state, the total
correlation cannot neatly divided into the quantum part and the classical part.
These two parts are much more ``entangled" with each other.

In this paper, we follow the lines of \cite{BGroisman05}. That is, roughly
speaking, the total correlation comes from the quantum part and the classical
one. Intuitively, the quantum correlation is more flimsy than the classical one,
and the classical correlation should be larger than the quantum correlation in
the mixed state. An obvious instance is that for the separable state, there is
no entanglement while classical correlation exists, in which case all the total
correlation comes from the classical part. Is it possible that the quantum
correlation is larger than the classical part? For this purpose, we would like
to adopt the quantum entanglement measure as large as possible. It is proved
that all the reasonable entanglement measures is not larger than the
entanglement of formation \cite{3H}. So, we take the entanglement of formation
as the quantum correlation and the classical correlation is defined as the total
correlation minus the quantum part. Before we discuss the main result, we argue
that the total correlation minus the entanglement of formation is the
non-erasable correlation under the constraint that entanglement is preserved.
Recall that the entanglement of formation is originally proposed to describe the
process of preparation of an entangled state under local operation and classical
communication (LOCC). From Eq. (\ref{eq:Ef}), we can see that the entanglement
of formation corresponds to a specified decomposition of the density matrix. In
experiment, to have a given decomposition, one of experimentalists, say Alice,
prepares the states $\{ |\psi_i\rangle\}$ according the probability distribution
$\{p_i\}$. Therefore, the actual state that describe the initial state of the
preparation process is
\begin{equation}
\bar\rho=\sum_i p_i |i\rangle\langle i |\otimes \ket{\psi_i}\bra{\psi_i},
\label{eq:varrho}
\end{equation}
where $\{|i\rangle\}$ are the flags, a set of orthogonal basis for Alice to
distinguish $|\psi_i\rangle$. After compression, Alice send the subsystem
through an ideal quantum channel to the another experimentalist Bob who need to
know which one he receives because in general he cannot decompress the state
without destroying the entanglement of the state. He requires the information to
distinguish $|\psi_i\rangle$. Therefore, the flags state are also needed to be
sent though this task does not require an ideal quantum channel. A classical
channel is enough. In order to obtain the goal state $\rho$ from the prepared
state $\bar\rho$, both Alice and Bob are required to erase the flag memory which
is used to store the information of the set of orthogonal basis. This procedure
decreases the classical correlation but preserves the entanglement of formation.
However any more information cannot be erased further or entanglement will be
destroyed. Therefore, the remaining part of the correlation represents the
non-erasable classical correlation between two subsystems $\rho$ under the
preservation of quantum correlation, and is calculated by
\begin{eqnarray}
S_C=S(1:2)-S_f. \label{eq:Sc}
\end{eqnarray}

In short, we argue that the mutual information is taken as the total
correlation, the entanglement of formation is taken as the quantum correlation,
and difference of them is the classical correlation in the meaning of
non-erasable classical correlation. Especially for an arbitrary two-qubit
system, the total, quantum and classical correlations can be easily calculated
and we can compare the quantum correlation and the classical one quantitatively.

\section{Environment's effects on correlations}
\label{sec:enviroment}

In this section, we use the Heisenberg dimer as a prototype model to show the
interesting behavior of the total, quantum and classical correlations under
different environment. The model Hamiltonian reads
\begin{eqnarray}
H= && J\left[{1-\gamma \over 2}(\sigma_1^x\sigma_2^x+\sigma_1^y\sigma_2^y)
+{1+\gamma \over 2}\sigma_1^z\sigma_2^z\right] \nonumber \\
&& +B_1\sigma_1^z+B_2\sigma_2^z, \label{eq:H}
\end{eqnarray}
where $\sigma_i^\alpha$ ($\alpha=x,y,z$) are the Pauli matrices, $J$ is the
strength of Heisenberg interaction, and $B_1,B_2$ are the external magnetic
fields. For simplicity, we choose $J$ as the energy unit. The parameter
$\gamma$, which ranges from $-1$ to 1, adjusts the anisotropic interactions.

\subsection{Anisotropic Heisenberg model}

We first consider the case of $B_1=B_2=0$. The eigenstates and eigenvalues are
\begin{eqnarray}
\ket{\psi_0} &=& {1 \over \sqrt{2}}(\ket{\uparrow \downarrow }-
\ket{\downarrow \uparrow }), \; E_0={-3+\gamma \over 2}; \nonumber \\
\ket{\psi_1} &=& {1 \over \sqrt{2}}(\ket{\uparrow \downarrow }
+\ket{\downarrow \uparrow }), \; E_1={1-3\gamma \over 2}; \nonumber \\
\ket{\psi_2} &=& \ket{\uparrow \uparrow }, \; E_2={1+\gamma \over 2}; \nonumber \\
\ket{\psi_3} &=& \ket{\downarrow \downarrow }, \; E_3={1+\gamma \over 2}.
\end{eqnarray}
The ground state is $\ket{\psi_0}$ for $\gamma \neq 1$. At the thermal
equilibrium, the density matrix of the system is
\begin{equation}
\rho(T)=\eta \left(
\begin{array}{cccc}
e^{-(1+\gamma) \over T} & 0 & 0 & 0 \\
0 & {\rm cosh}{1-\gamma \over T} & -{\rm sinh}{1-\gamma \over T} & 0 \\
0 & -{\rm sinh}{1-\gamma \over T} & {\rm cosh}{1-\gamma \over T} & 0 \\
0 & 0 & 0 & e^{-(1+\gamma) \over T}
\end{array}
\right), \label{eq:rho1}
\end{equation}
where the Boltzmann's constant $k_B$ is set as one and
\begin{equation}
\eta = {1 \over 2\left[{\rm cosh}{1-\gamma \over T} +
e^{-(1+\gamma)/T}\right]}. \nonumber
\end{equation}
\begin{figure}
\includegraphics[width=4cm]{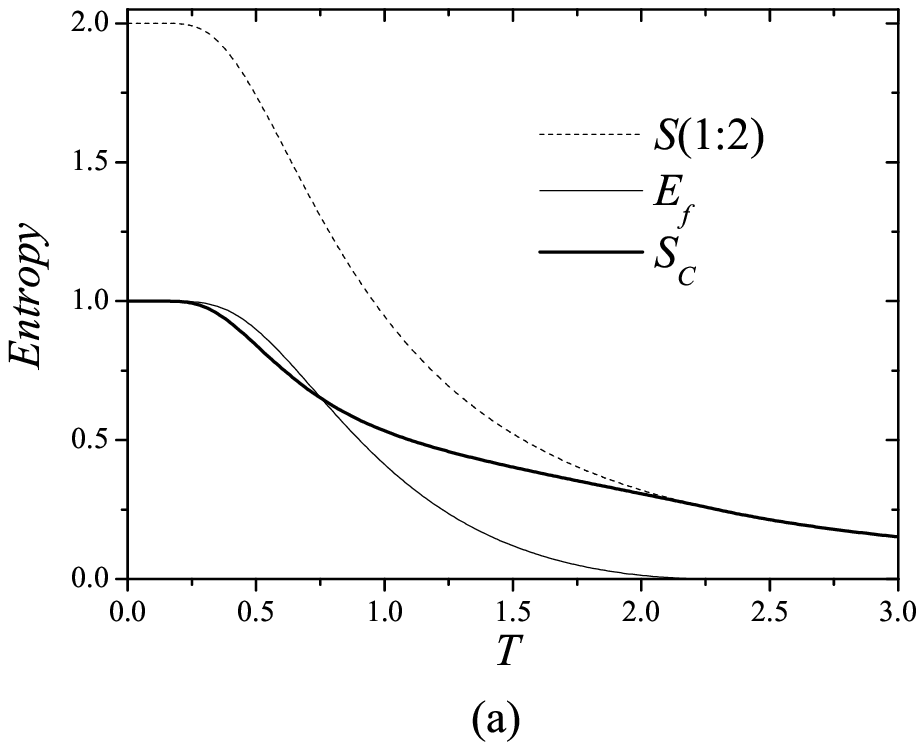}
\includegraphics[width=4cm]{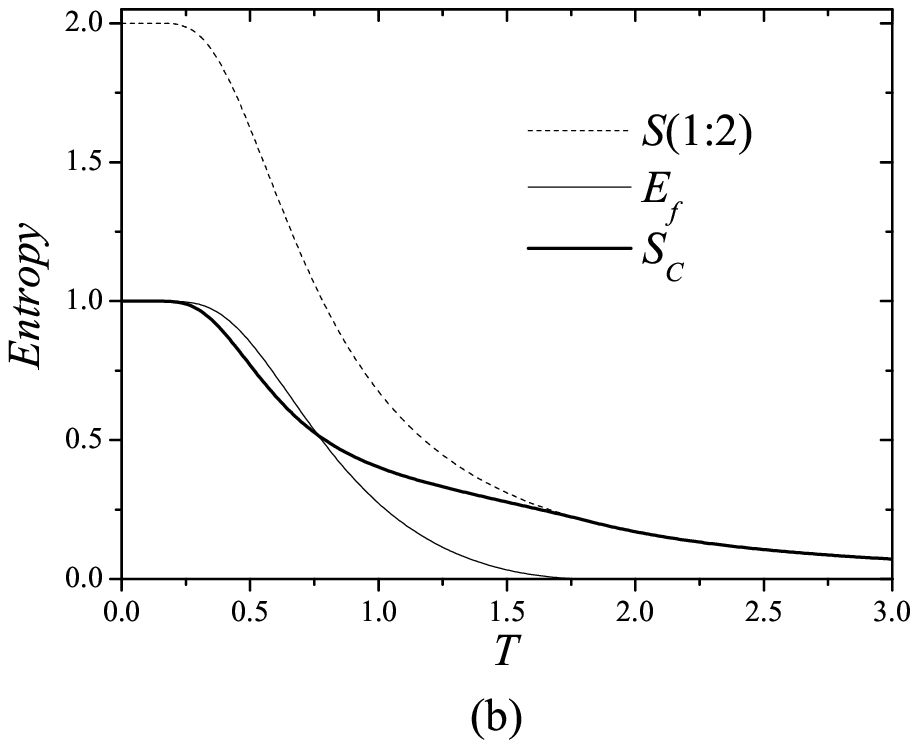}
\includegraphics[width=4cm]{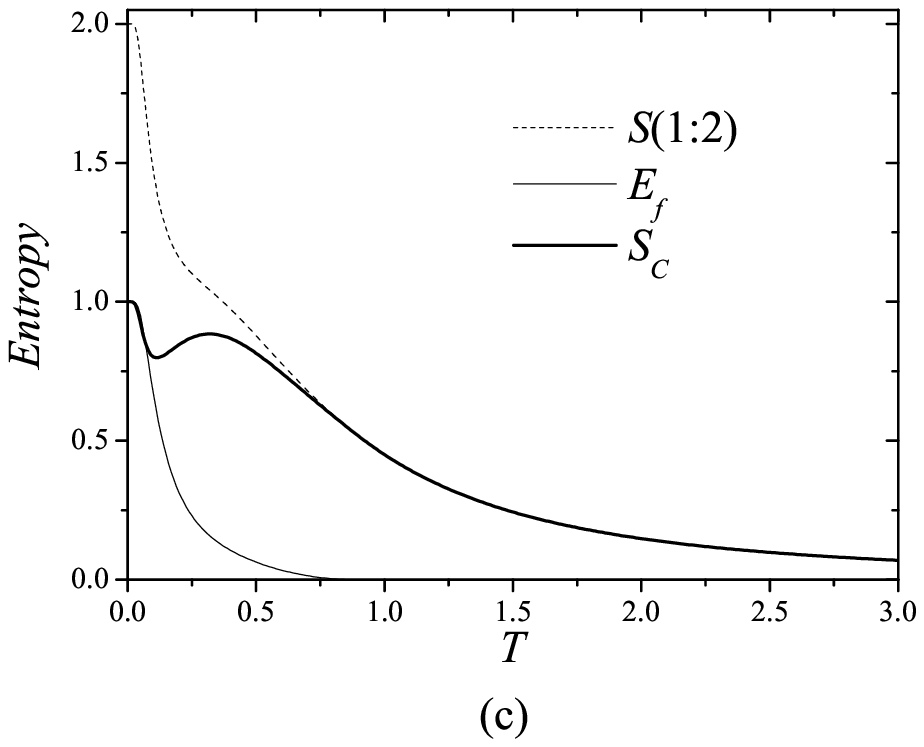}
\includegraphics[width=4cm]{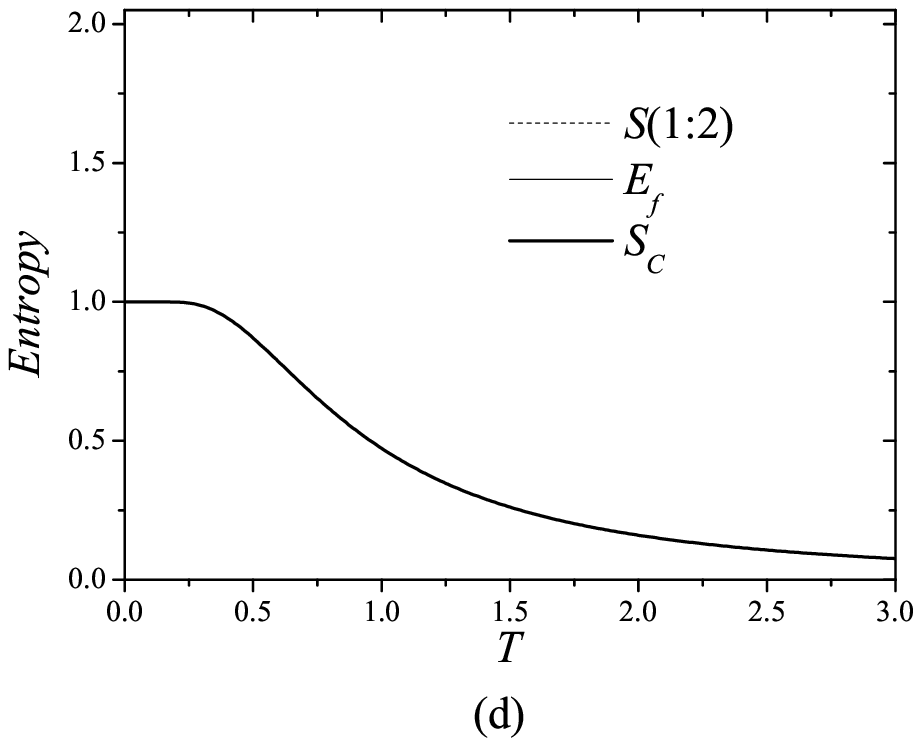}
\caption{\label{fig:anisotropic}The total correlation $S(1:2)$,
quantum correlation $E_f$ and classical correlation $S_C$ versus
temperature $T$, in the anisotrpic Heisenberg dimer with differnt
values of parameter $\gamma$. In the Fig. (d), the $E_f=0$ and the
curves of $S(1:2)$ and $S_C$ are overlap. (a) $\gamma=-1$, (b)
$\gamma=0$, (c) $\gamma=0.9$ and (d) $\gamma=1$. }
\end{figure}
From the density matrix (\ref{eq:rho1}), the total, quantum and classical
correlations can be calculated directly. The results is shown in Fig.
\ref{fig:anisotropic}. We see several interesting features from this figure.
First, at high temperature, all correlations approach zero. This is because that
the occupation probabilities of the unentangled states will be enhanced and the
correlations will be diluted. The thermal fluctuation is the leading effects.
Second, at certain temperature range, the classical correlation exceeds the
quantum correlation. This is obvious in cases of small $\gamma$ (Fig.
\ref{fig:anisotropic}(a) and (b)). Third, when $\gamma$ is close to 1 (Fig.
\ref{fig:anisotropic}(c)), the classical correlation may exhibit a local minimum
at low temperature. It is worth noting that the quantum correlation is smaller
for a larger $\gamma$. The physical interpretation is that a larger $\gamma$
corresponds to a more classical model, hence less amount of entanglement. When
$\gamma$ reaches 1, the quantum correlation vanishes for all temperature
(Fig.~\ref{fig:anisotropic}(d)). This is expected because when $\gamma=1$ all
eigenstates of the Hamiltonian are unentangled states, hence the resulting
density matrix $\rho(T)$ is separable for all temperature.

\begin{figure}
\includegraphics[width=7cm]{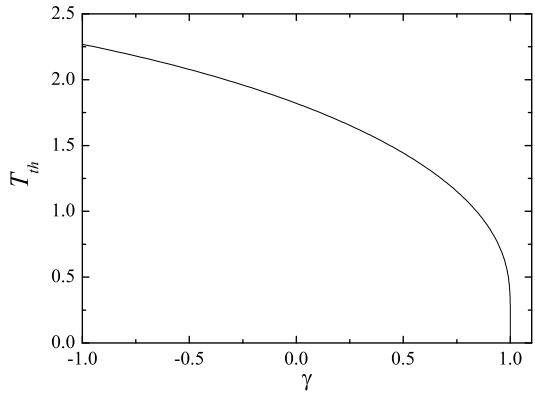}
\caption{\label{fig:thre1}The threshold temperature $T_{\rm th}$ versus
anisotropic parameter $\gamma$ in the Heisenberg dimer. The quantum correlation
of the system vanishes if $T>T_{\rm th}$.}
\end{figure}

The above features can be illustrated more clearly by defining a threshold
temperature $T_{\rm th}$, which is a function of $\gamma$. Above the $T_{\rm
th}$, the quantum correlation completely vanishes. The quantum correlation
reaches zero when the concurrence reaches zero. From
Eqs.~(\ref{eq:con}),(\ref{eq:varrho}) and (\ref{eq:rho1}), the concurrence of
the system is
\begin{equation}
C={\rm max} \left\{ {{\rm sinh}{1-\gamma \over T}-e^{-(1+\gamma)/T}
\over {\rm cosh}{1-\gamma \over T}+e^{-(1+\gamma)/T}},0 \right\}. \nonumber
\end{equation}
The $C=0$ requires ${\rm sinh}{1-\gamma \over T} \leq e^{-(1+\gamma)/T}$. Then
the threshold temperature $T_{\rm th}$ should satisfy
\begin{equation}
\gamma={T_{\rm th} \over 2}{\rm ln}\left( e^{2/T_{\rm th}}-2 \right).
\end{equation}
The plot of $T_{\rm th}$ versus $\gamma$ is shown in Fig. \ref{fig:thre1}. It is
obvious that $T_{\rm th}$ drops when $\gamma$ increases. From
Fig.\ref{fig:thre1}, we can divide the whole plane into two regions. Below the
line of $T_{\rm th}$, the system has both quantum and classical correlations,
while above the line, the system has only classical correlation.

\begin{figure}
\includegraphics[width=4cm]{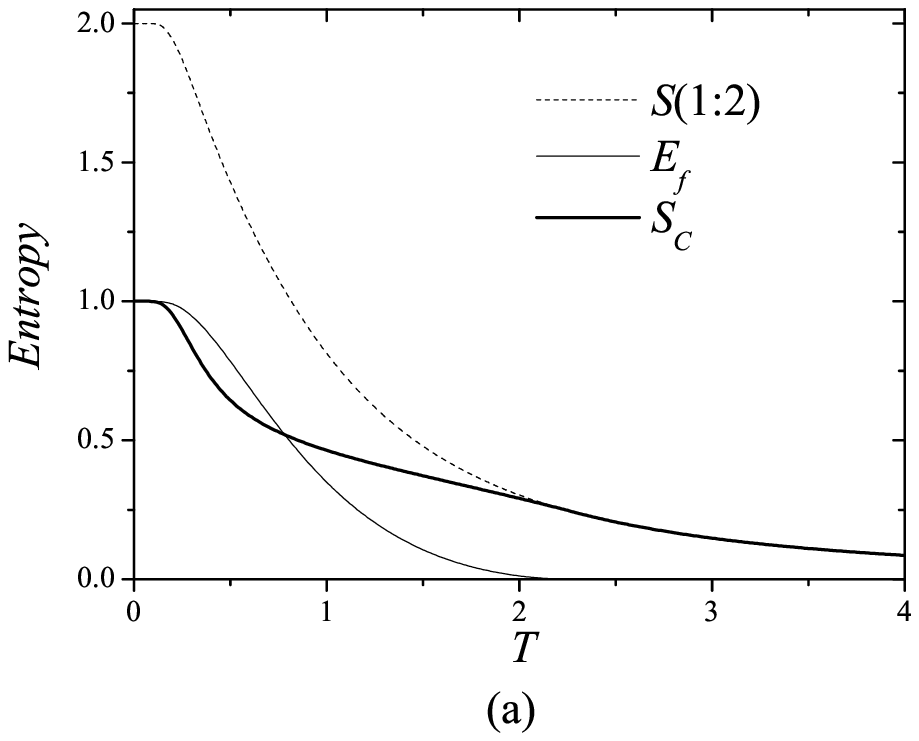}
\includegraphics[width=4cm]{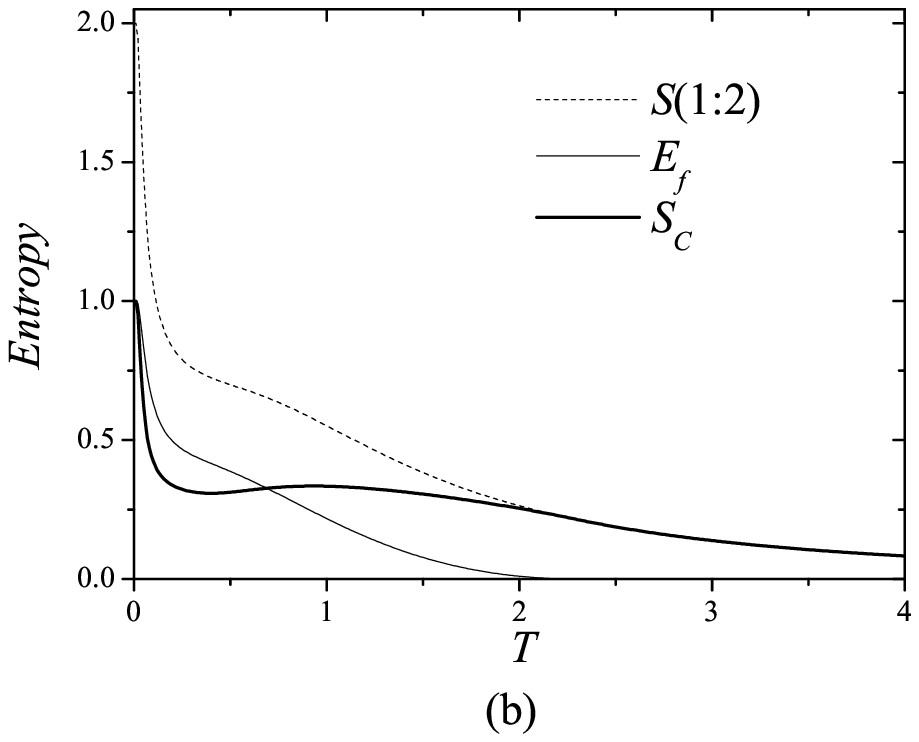}
\includegraphics[width=4cm]{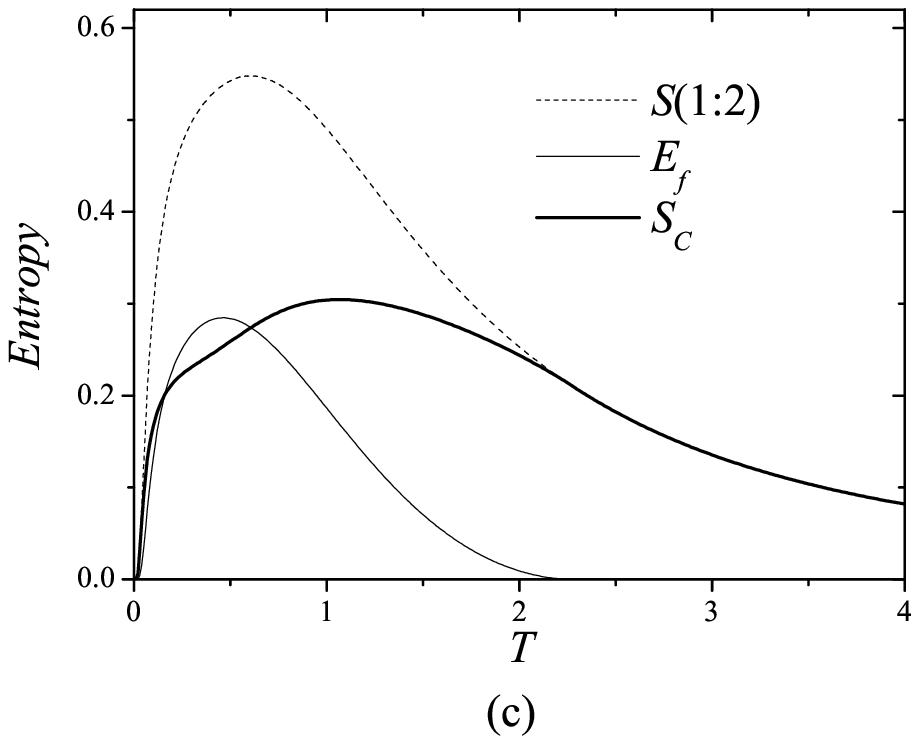}
\includegraphics[width=4cm]{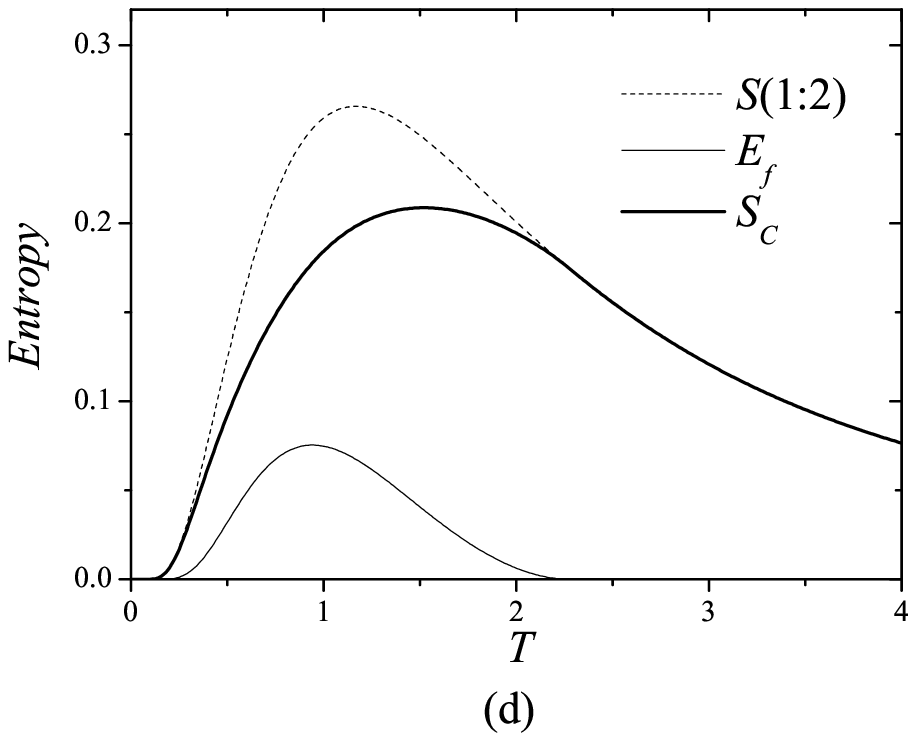}
\caption{\label{fig:XY_1} The total correlation $S(1:2)$, quantum correlation
$E_f$ and classical correlation $S_C$ versus temperature $T$, in the $XY$ model
under a uniform magnetic fields $B_1=B_2$. Varying the strength of the fields,
four typical cases are shown: (a) $B_1=B_2=0.5$, (b) $B_1=B_2=0.95$, (c)
$B_1=B_2=1.05$ and (d) $B_1=B_2=1.5$.}
\end{figure}

\begin{figure}
\includegraphics[width=4cm]{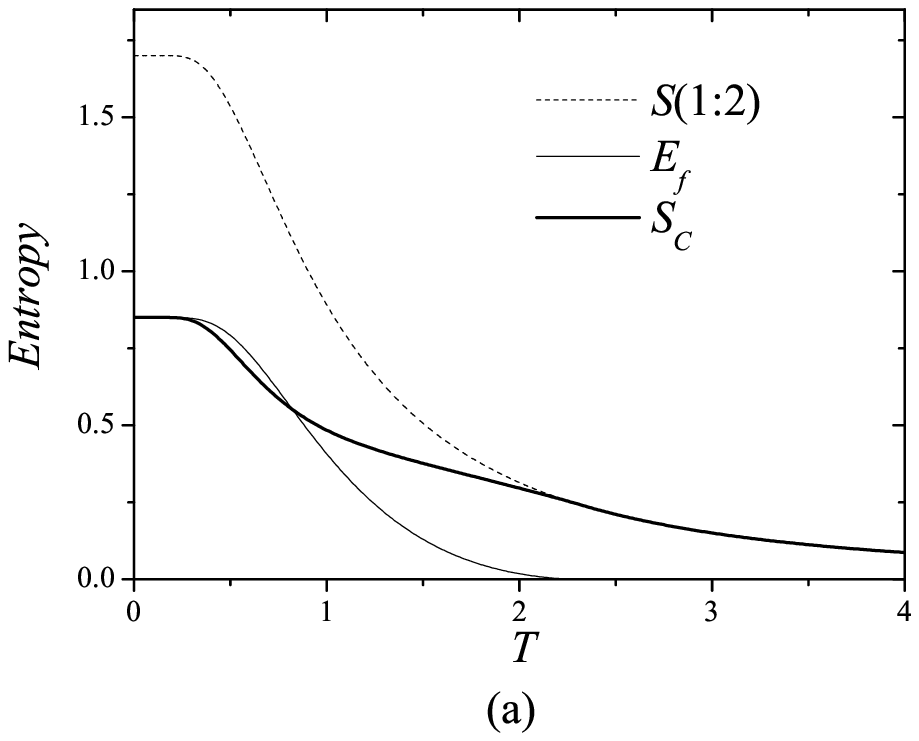}
\includegraphics[width=4cm]{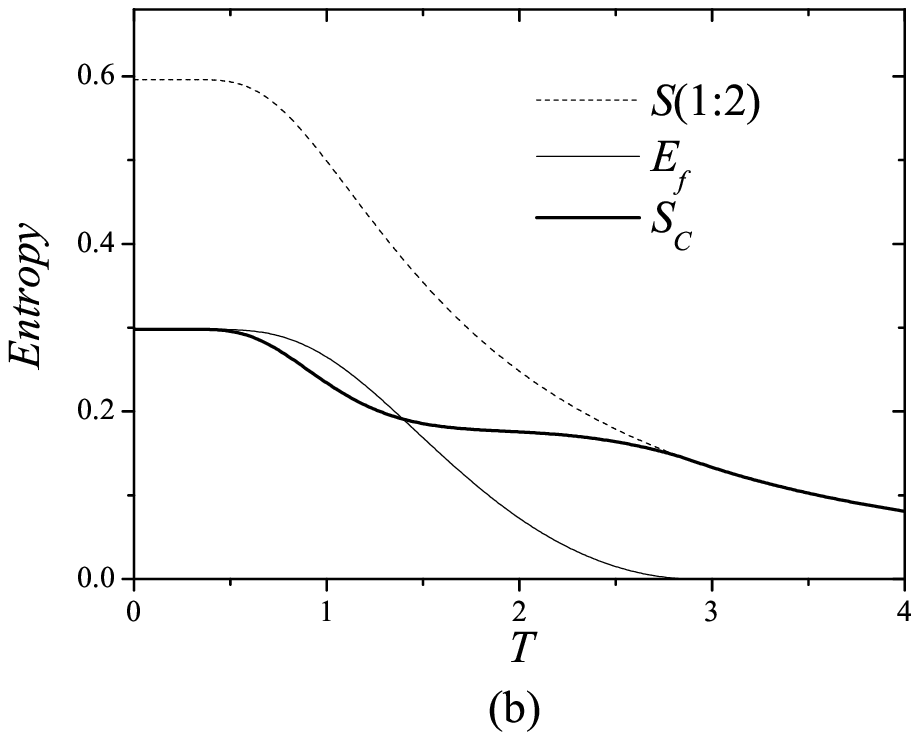}
\caption{\label{fig:XY_2}The total correlation $S(1:2)$, quantum correlation
$E_f$ and classical correlation $S_C$ versus temperature $T$, in the $XY$ model
under nonuniform field: (a) $B_1=0.5, B_2=-0.5$ and (b) $B_1=2, B_2=-2$. }
\end{figure}

\subsection{XY model with nonuniform magnetic field}

\begin{figure*}
\begin{tabular}{|@{\hspace{4mm}}c@{\hspace{4mm}}|@{\hspace{4mm}}c@{\hspace{4mm}}|@{\hspace{4mm}}c@{\hspace{4mm}}|}
\includegraphics[width=4cm]{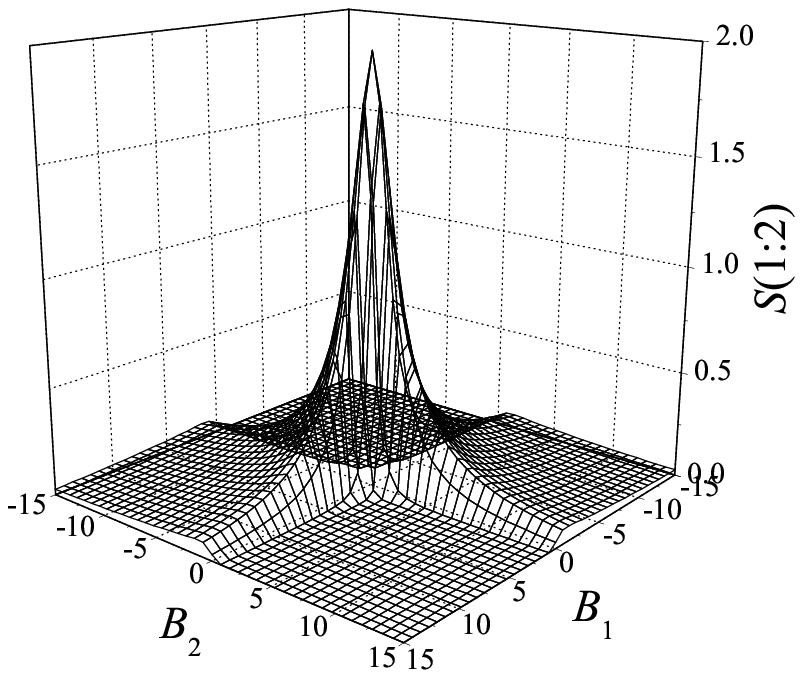} &
\includegraphics[width=4cm]{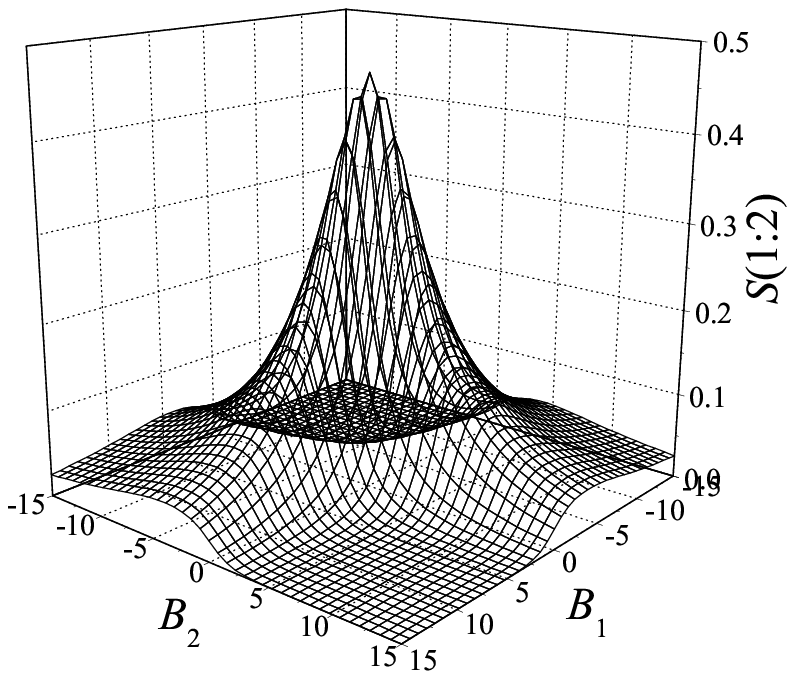} &
\includegraphics[width=4cm]{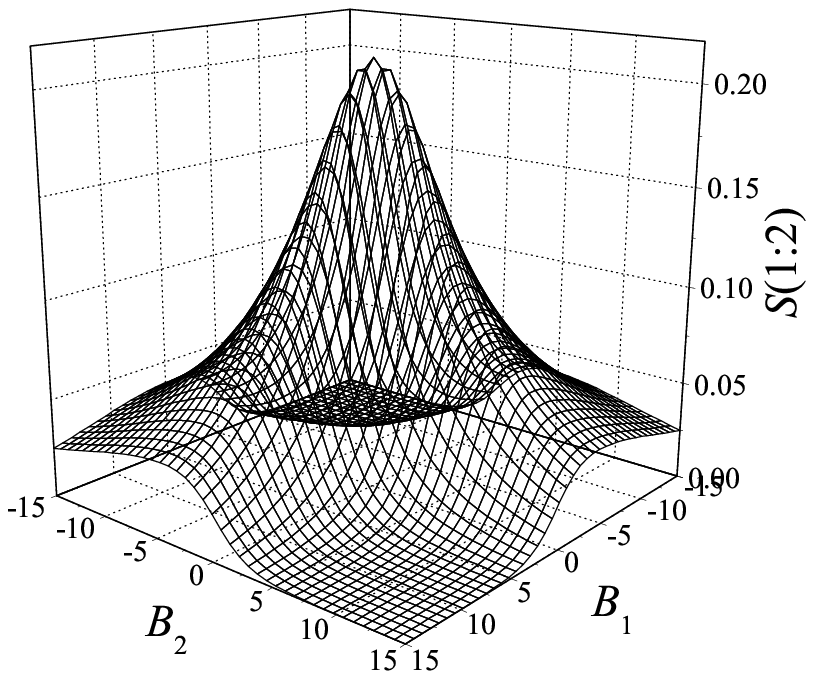} \\

\includegraphics[width=4cm]{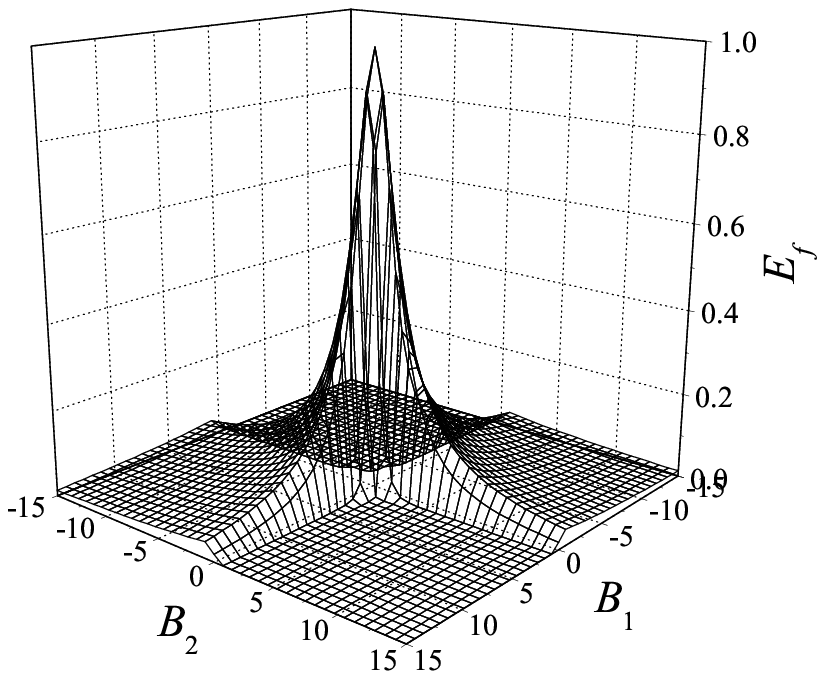} &
\includegraphics[width=4cm]{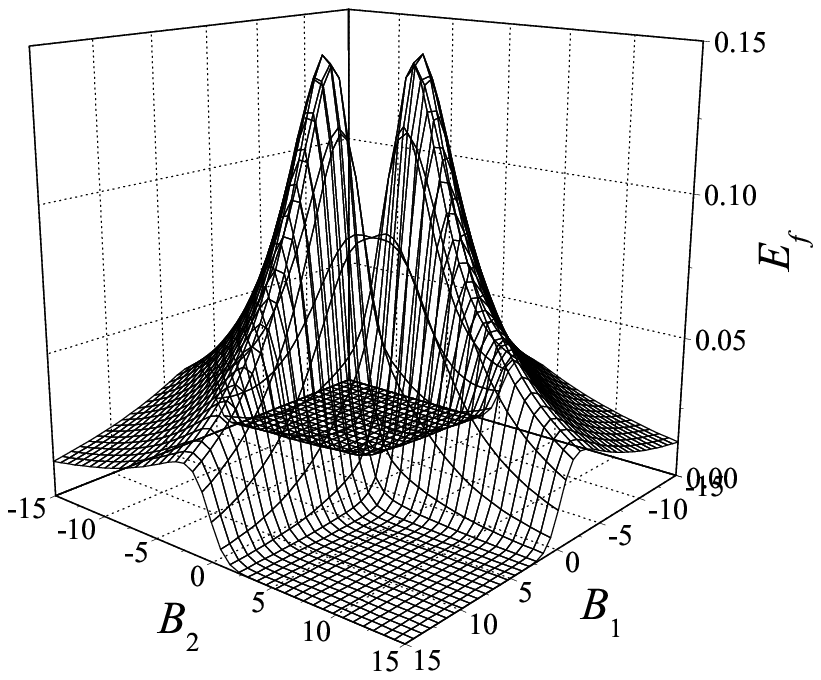} &
\includegraphics[width=4cm]{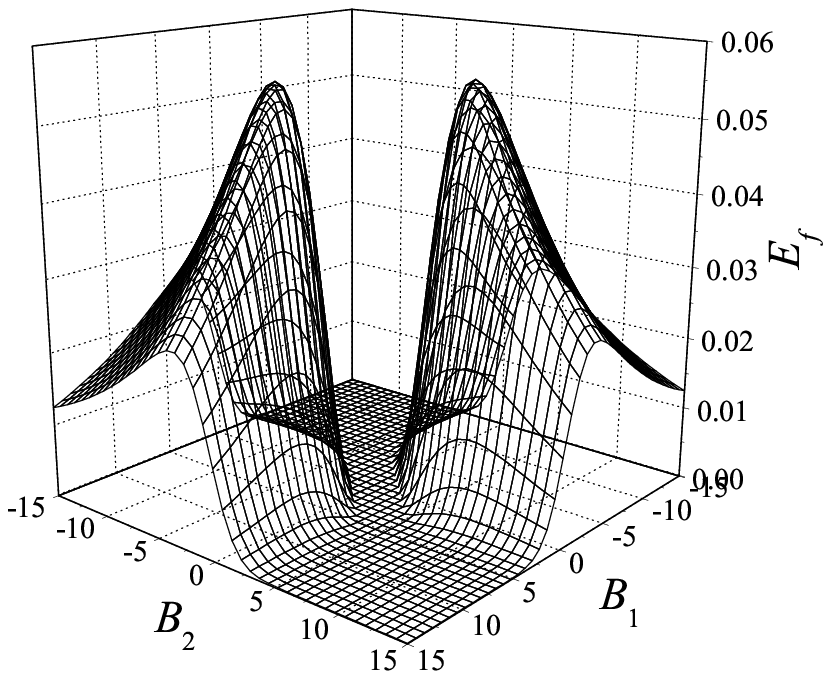} \\

\includegraphics[width=4cm]{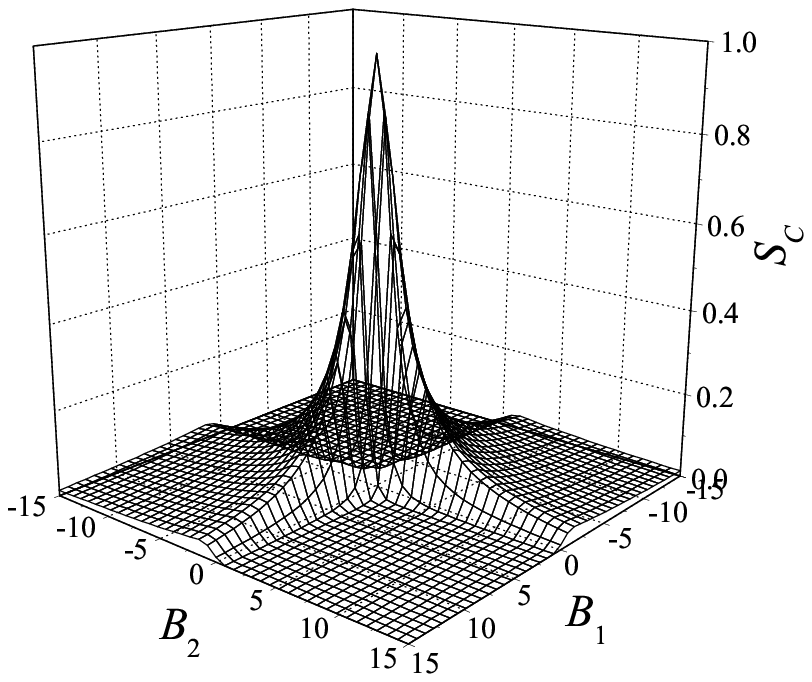} &
\includegraphics[width=4cm]{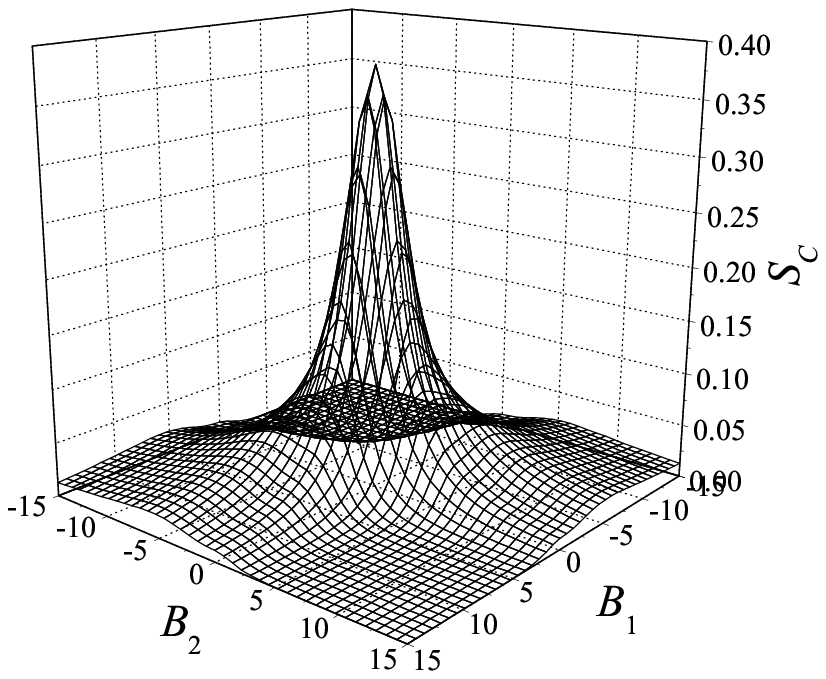} &
\includegraphics[width=4cm]{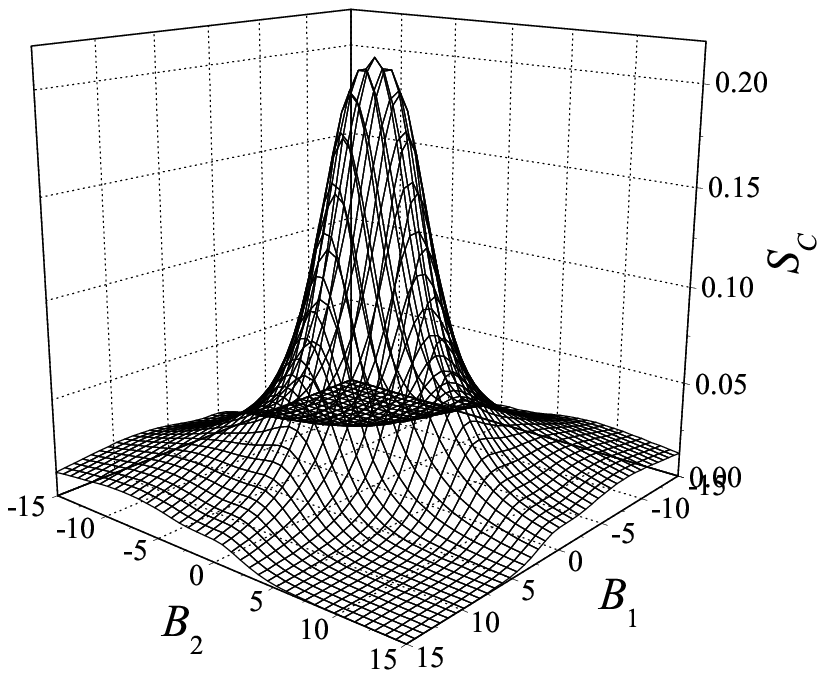} \\[2mm]
(a) & (b) & (c)
\end{tabular}
\caption{\label{fig:XY_3}(From top to bottm) the total correlation $S(1:2)$,
quantum correlation $E_f$ and classical correlation $S_C$ versus the external
magnetic fields $B_1$ and $B_2$ at different fixed temperatures $T$ in the $XY$
model. Three typical cases are shown in different columns: (a) $T=0.3$, (b)
$T=1.6$ and (c) $T=2.5$.}
\end{figure*}

Now we investigate the correlation effects of the external magnetic fields. We
only consider the case $\gamma=-1$ and other $\gamma$ can be obtained similarly.
Then the eigenstates and eigenvalues of the Hamiltonian are
\begin{eqnarray}
&& \ket{\Psi_1} = \ket{\uparrow \uparrow }, \; E_1=B_1+B_2; \nonumber \\
&& \ket{\Psi_2} = \ket{\downarrow \downarrow }, \; E_2=-(B_1+B_2); \nonumber \\
&& \ket{\Psi_\pm} = {1 \over N_\pm}\left[ {(B_1-B_2)\pm \sqrt{D} \over 2}
\ket{\uparrow \downarrow}+\ket{\downarrow \uparrow}\right], \nonumber \\
&& E_\pm = \pm \sqrt{D},
\end{eqnarray}
where $D=(B_1-B_2)^2+4$ and $N_\pm$ are the normalization factors. The thermal
equilibrium state can be described by the density matrix
\begin{equation}
\rho(T)={1 \over Z}\left(
\begin{array}{cccc}
d & 0 & 0 & 0 \\
0 & b-c & -s & 0 \\
0 & -s & b+c & 0 \\
0 & 0 & 0 & d^{-1}
\end{array}
\right), \label{eq:rho2}
\end{equation}
where $Z=2\{{\rm cosh}{[(B_1+B_2)/T]} + {\rm cosh}{(\sqrt{D}/ T)}\}$,  $b={\rm
cosh}(\sqrt{D}/T)$, $c={\rm sinh}{(\sqrt{D} /T)}(B_1-B_2)/\sqrt{D}$, $s={2 }{\rm
sinh}{(\sqrt{D} / T)}/\sqrt{D}$ and $d=\exp{[-(B_1+B_2)/T]}$.

We first study the correlations under uniform magnetic fields at finite
temperatures. The results are shown in Fig. \ref{fig:XY_1} (a-d). Clearly, if
$B_1$ is small, the ground state  $\ket{\Psi_-}$ is a superposition of two
antiferromagnetic basis, and is entangled. If the system is subject to a thermal
environment, the contribution from the other eigenstates (two of them are
separable) will suppress both the quantum and classical correlation. However, if
$B_1$ is large enough, the ground state becomes $\ket{\Psi_2}$, which is fully
polarized and not entangled. The classical correlation, whose value is equal to
the quantum one for a pure state, is also zero. The thermal fluctuation, as can
see from Fig. \ref{fig:XY_1} (c and d), increases both the quantum and classical
correlations at low temperatures. Moreover, an interesting observation is that
there exists a range where the quantum correlation exceeds the classical one. In
addition, the larger the external field, the smaller the quantum correlation. It
is because a large-field setting corresponds to a more classical model. It may
be interesting to note that the threshold temperature of the quantum correlation
is independent of the field \cite{YSun03}. All three correlations approach or
equal zero at high temperatures.

If the directions of two external fields are opposite to each other and the
strength are the same, we find that all the total, classical and quantum
correlations show comparatively gentle changes against the temperature and
fields (Fig.~\ref{fig:XY_2} (a) and (b)). The figure is not difficult to
interpret, as we argued for the case of $B_1=B_2$. The main difference is that
the a larger $B_1$ here may leads to a higher threshold temperature $T_{\rm
th}$. In order to see the role of the nonuniform field, we show three
correlations against fields at some fixed temperatures in Fig.~\ref{fig:XY_3}
(The results of quantum correlation, which have already been obtained by Sun
{\it etal} \cite{YSun03}, are also presented for comparison).  At low
temperature, we notice that the three correlations are sharply peaked at zero
fields. They decay rapidly with the increasing fields if the fields have the
same direction, while decay comparatively slowly if the fields have opposite
directions. This means that the correlation effects can be adjusted by the
uniform fields. At some higher temperature, the peak of the quantum correlation
splits into two in the region $B_1B_2<0$ (Fig. \ref{fig:XY_3}(b)). Therefore,
the nonuniform fields may enhance the quantum correlation, while the uniform
fields always destructs it. At very high temperature, the peaks are completely
separated as shown in Fig.~\ref{fig:XY_3}(c). A region with zero quantum
correlation appears between the peaks. This implies that the nonuniform field
can be used as a switch to turn on and off the quantum correlation
\cite{YSun03}. Meanwhile, unlike the quantum correlation, the classical
correlation always decreases with the increasing external magnetic fields. Which
means that the external magnetic fields have different effects to the quantum
correlation and to the classical correlation.

\section{Discussions and summary}
\label{sec:sum}

In this paper, we provide quantification of the total, quantum and classical
correlations in a general bipartite system. In order to see their properties in
a realistic system, we study them in an anisotropic Heisenberg model at finite
temperatures. We find that the quantum correlation always decreases with the
increasing temperature, while the classical one may increase in some temperature
range. More interestingly, the classical correlation is not always larger than
the quantum one, which actually is beyond the general physical intuition
\cite{BGroisman05}. We also investigate the three correlations in the XY model
under a nonuniform magnetic field. We find that the fields may enhance the
quantum correlation, which is very different from the effect of fields to the
classical correlation. In short, our results imply that the environmental
parameters (temperature, magnetic fields) demonstrate obviously different
effects to the quantum correlation and the classical correlation.

This work is partially supported by Direct Grant of CUHK (A/C 2060286), the
Earmarked Grant for Research from the Research Grants Council of HKSAR, China
(Project CUHK N\_CUHK204/05), and NSFC under Grant No. 10574150. D. Yang
acknowledges the financial support from the C. N. Yang Foundation. J. P. Cao and
D. Yang are grateful for the hospitality of the Department of Physics at CUHK.


\begin{references}

\bibitem{EPR}
A.~Einstein, B.~Podolsky and N.~Rosen, Phys.~Rev. {\bf 47}, 777 (1935).

\bibitem{Nielsen1}
M. A. Nilesen and I. L. Chuang, {\it Quantum Computation and Quantum
Information} (Cambridge University Press, Cambridge, England, 2000)

\bibitem{AGalindo02}
A. Galindo and M. A. Martin-Delgado, Rev. Mod. Phys. {\bf 74}, 347 (2002).

\bibitem{Bennett}
C. H. Bennett and S. J. Wiesner, Phys. Rev. Lett., {\bf 68}, 557 (1992); C. H.
Bennett, G. Brassard, C. Crepeau, R. Jozsa, A. Peres, and W. Wootters, Phys.
Rev. Lett., {\bf 70}, 1895 (1993); A. K. Ekert, J. G. Rarity, P. R. Tapster, and
G. M. Palma, Phys. Rev. Lett. {\bf 69}, 1293 (1992).


\bibitem{XWang01}
X.~Wang, Phys. Rev. A {\bf 64}, 012313 (2001); X. Wang,  Phys. Rev. A {\bf 66},
044305 (2002).

\bibitem{PZanardi}
P. Zanardi, X. Wang, J. Phys. A: Math. and Gen. {\bf 35} 7947 (2002).

\bibitem{XQXi02}
X. Q. Xi, S. R. Hao, W. X. Chen, and R. H. Yue, Chin. Phys. Lett. {\bf 19}, 1044
(2002).

\bibitem{CAnteneodo03}
C. Anteneodo and A. M. C. Souza, J. Opt. B: Quantum Semiclass. Opt. {\bf 5} 73
(2003).

\bibitem{YSun03}
Y. Sun, Y. G. Chen and H. Chen, Phys. Rev. A {\bf 68}, 044301 (2003).




\bibitem{HV}
L. Henderson, V. Vedral, quant-ph/0105028

\bibitem{BGroisman05}
B. Groisman, S. Popescu and A. Winter, Phys. Rev. A {\bf 72}, 032317 (2005).


\bibitem{PV}
M. B. Plenio, S. Virmani, Quant. Inf. Comp. {\bf 7}, 1 (2007), quant-ph/0504163


\bibitem{CHBennett96}
C. H. Bennett, D. P. DiVincenzo, J.~Smolin, and W.~K.~Wooters, Phys. Rev. A {\bf
54}, 3824 (1996).


\bibitem{SHill97}
S. Hill and W. K. Wootters, Phys.~Rev.~Lett. {\bf 78}, 5022 (1997); W. K.
Wootters, Phys.~Rev.~Lett. {\bf 80}, 2245 (1998).

\bibitem{AWehrl94}
A.~Wehrl, Rev.~Mod.~Phys. {\bf 66}, 129 (1994).




\bibitem{3H}
M. Horodecki, P. Horodecki, R. Horodecki, Phys. Rev. Lett. {\bf 84}, 2014 (2000)





\end{references}
\end{document}